\documentclass[sigconf]{acmart} 
\usepackage{csquotes}
\usepackage{amsfonts,amsmath,amsthm}

\usepackage{float} 
\usepackage{multirow} 
\usepackage{makecell}
\usepackage{enumitem}
\usepackage{dsfont}
\usepackage{tabularx}
\usepackage{xcolor}
\usepackage{placeins}
\newenvironment{compact_enum}{
  \begin{itemize}[leftmargin=*, itemsep=0pt, parsep=2pt, topsep=2pt]
}{
  \end{itemize}
}

\AtBeginDocument{%
  }

\copyrightyear{2026}
\acmYear{2026}
\setcopyright{cc}
\setcctype{by}
\acmConference[KDD '26]{Proceedings of the 32nd ACM SIGKDD Conference on Knowledge Discovery and Data Mining V.2}{August 09--13, 2026}{Jeju Island, Republic of Korea}
\acmBooktitle{Proceedings of the 32nd ACM SIGKDD Conference on Knowledge Discovery and Data Mining V.2 (KDD '26), August 09--13, 2026, Jeju Island, Republic of Korea}
\acmDOI{10.1145/3770855.3818476}
\acmISBN{979-8-4007-2259-2/2026/08}

\begin{document}

\title{SAGE: Scalable AI Governance \& Evaluation}


\settopmatter{authorsperrow=3}

\newcommand{\linkedinAffil}{%
  \affiliation{%
    \institution{LinkedIn Corporation}%
    \city{Mountain View}%
    \state{CA}%
    \country{USA}%
  }%
}

\author{Benjamin H. Le}
\email{ble@linkedin.com}
\author{Xueying Lu}
\email{xueylu@linkedin.com}
\author{Nicholas Stern}
\email{nstern@linkedin.com}
\author{Wenqiong Liu}
\email{ecliu@linkedin.com}
\author{Igor Lapchuk}
\author{Xiang Li}
\authornote{Work done while at LinkedIn.}
\author{Baofen Zheng}
\linkedinAffil

\author{Kevin Rosenberg}
\author{Jiewen Huang}
\author{Zhe Zhang}
\author{Abraham Cabangbang}
\author{Satej Milind Wagle}
\author{Jianqiang Shen}
\author{Raghavan Muthuregunathan}
\linkedinAffil

\author{Abhinav Gupta}
\authornotemark[1]
\author{Mathew Teoh}
\authornotemark[1]
\author{Andrew J. N. Kirk}
\author{Thomas Kwan}
\author{Jingwei Wu}
\author{Wenjing Zhang}
\linkedinAffil

\renewcommand{\shortauthors}{Benjamin H. Le et al.}


\begin{abstract}
Evaluating relevance in large-scale search systems is fundamentally constrained by the governance gap between nuanced, resource-constrained human oversight and the high-throughput requirements of production systems. While traditional approaches rely on engagement proxies or sparse manual review, these methods often fail to capture the full scope of high-impact relevance failures. We present \textbf{SAGE} (Scalable AI Governance \& Evaluation), a framework that operationalizes high-quality human product judgment as a scalable evaluation signal. At the core of SAGE is a bidirectional calibration loop where natural-language \emph{Policy}, curated \emph{Precedent}, and an \emph{LLM Surrogate Judge} co-evolve. SAGE systematically resolves semantic ambiguities and misalignments, transforming subjective relevance judgment into an executable, multi-dimensional rubric with near human-level agreement. To bridge the gap between frontier model reasoning and industrial-scale inference, we apply teacher–student distillation to transfer high-fidelity judgments into compact student surrogates at \textbf{92$\times$} lower cost. Deployed within LinkedIn Search ecosystems, SAGE guided model iteration through simulation-driven development, distilling policy-aligned models for online serving and enabling rapid offline evaluation. In production, it powered policy oversight that measured ramped model variants and detected regressions invisible to engagement metrics. Collectively, these drove a \textbf{0.25\%} lift in LinkedIn daily active users.
\end{abstract}  




\maketitle

\section{Introduction}
Industrial search systems are shifting from keyword-based retrieval toward semantic models that infer user intent from natural language \cite{shah2025towards, ye2025applying}. In large-scale platforms such as LinkedIn Job Search and People Search, members routinely express multifaceted intent that combines role titles, seniority constraints, skills, organizational attributes, and professional relationships \cite{kenthapadi2017personalized}. While semantic models substantially expand expressiveness and recall, they also complicate the problem of relevance governance: specifying, applying, and enforcing what the system \emph{ought} to return across a vast and continuously evolving query–document space.

Product owners possess nuanced mental models of relevance quality, yet these latent judgments are difficult to operationalize at the scale and consistency required by modern AI systems. This governance gap leaves development teams without a reliable, human-centric benchmark for model iteration and release gating. Current relevance governance typically relies on suboptimal proxies: manual editorial review provides high-fidelity signal but fails to cover the combinatorial diversity of natural-language query spaces, while engagement metrics introduce delayed feedback loops and systematic biases \cite{joachims2017unbiased}, such as popularity amplification, that often diverge from intended relevance. Consequently, these proxies often fail to capture the full scope of high-impact relevance failures.

Recent advances in large language models (LLMs) enable scalable evaluation by approximating expert judgment through instruction-following and reasoning \cite{zheng2023judging, gu2024survey}. However, generic LLM-as-a-judge approaches often encode implicit assumptions and suffer from domain mismatch, functioning as black-box graders rather than faithful surrogates of human product judgment \cite{wang2023llmfair, li2025curse}.

We introduce \textbf{SAGE} (Scalable AI Governance \& Evaluation), a framework that operationalizes human product judgment as a scalable, formal evaluation signal for industrial search. SAGE formalizes judgment through a natural-language \textit{Policy} specifying acceptable system behavior, a curated \textit{Precedent} grounding policy interpretation, and an \textit{LLM Surrogate Judge} calibrated to apply policy-consistent decisions. To support evaluation at scale, SAGE employs distillation to compress frontier LLM reasoning into efficient student surrogates, enabling both large-scale offline evaluation ($>10^7$ annotations per day) and low-latency online evaluation ($>10^4$ QPS).

We validate SAGE through its deployment in LinkedIn Search. SAGE guided model iteration through simulation-driven development, distilling policy-aligned models for online serving and enabling rapid offline evaluation. In production, it powered policy oversight that measured ramped model variants and detected regressions invisible to engagement metrics. Collectively these capabilities drove a \textbf{0.25\%} lift in LinkedIn daily active users. Our results demonstrate that human-centric governance can be formalized at scale to meet the demands of industry AI systems and steer model development toward member outcomes.

We make the following contributions to industry-scale human-centric evaluation: (1) \textbf{Bidirectional Calibration for Intent Alignment} that transforms heterogeneous human product judgment into versioned evaluation criteria. Unlike static LLM-based judges, it systematically surfaces policy ambiguities and judge misalignment to enable evaluation to co-evolve with human judgment. (2) \textbf{Interpretable Multi-Attribute Relevance Evaluation} that decomposes relevance into orthogonal attributes (e.g., Title, Seniority) to eliminate black-box opacity, enabling targeted failure attribution during bidirectional calibration and achieving near human-level agreement. (3) \textbf{High-Throughput Evaluation via Distillation} that compresses frontier LLM reasoning into student surrogates, achieving high fidelity while enabling large-scale offline evaluation and low-latency online evaluation.

The remainder of this paper is organized as follows. Section~\ref{sec:related_work} reviews related work on AI governance and evaluation. Section~\ref{sec:framework} formalizes the relevance governance problem and introduces the SAGE framework. Section~\ref{sec:teacher-judge} describes the design of the Teacher Judge and the bidirectional calibration loop used to align evaluation with product judgment. Section~\ref{sec:student-judge} presents the optimization techniques used to derive a cost-efficient Student Judge suitable for production-scale inference. Section~\ref{sec:use-cases} demonstrates using the Student Judge to drive production impact through distillation for online serving, offline candidate selection, online experimentation, and continuous relevance monitoring. Finally, Section~\ref{sec:conclusion} summarizes the results and key deployment lessons.

\section{Related Work}
\label{sec:related_work}
Industrial search systems have historically combined editorial relevance judgments with engagement signals to develop and validate ranking changes~\cite{Yin2016Ranking}. While essential for measuring user impact, engagement signals are slow to reveal subtle but high-impact relevance failures, and are difficult to apply consistently across the long tail of semantic, open-ended queries. Crowdsourcing human labeling can increase coverage, but inconsistency and annotator-bias distort the resulting ``ground truth''~\cite{geva2019annotator}. SAGE addresses this by concentrating human review on establishing precedent rather than exhaustively labeling data.

Recent work shows that LLMs can act as relevance and preference judges for machine learning systems~\cite{Mehrdad2024LLM, Thomas2023LLM, Zhang2024Explanations, hu2025training, lu2025vlm}. Beyond direct grading, rubric-based prompting decomposes relevance into explicit dimensions, improving clarity and enabling more controllable judgments~\cite{ye2023flask, Farzi2025Criteria, kim2024prometheus}. Open-source evaluation frameworks such as Ragas~\cite{es2023ragas} and G-Eval~\cite{liu2023geval} operationalize these patterns with general-purpose metrics for faithfulness, coherence, and answer relevance. However, these frameworks treat evaluation criteria as static inputs and assume a universal notion of quality. SAGE's relevance decomposability introduces structured, multi-attribute judgments, which is conceptually aligned with multi-dimensional evaluation frameworks such as FLASK~\cite{ye2023flask}. This multi-attribute decomposition, rather than a single scalar score, reduces the risk that stylistic factors (e.g., verbosity or confident tone) dominate the evaluation signal. In addition, SAGE transforms heterogeneous human product judgment into versioned evaluation criteria through bidirectional calibration, enabling evaluation to co-evolve with human judgment.

LLM-as-a-judge methods can exhibit systematic biases and sensitivity to prompt and presentation choices, motivating meta evaluation practices~\cite{zheng2023judging, wang2023llmfair}. \citet{Braun2025Scalable} explores mitigations such as structured judging protocols and aggregation across multiple evaluations to improve robustness; \citet{xieatc} focuses on aggregating heterogeneous annotator judgments through pairwise comparisons. Our work builds on these findings while adopting different design principles by treating calibration to product owner precedent as a first-class requirement: evaluators should be verified as consistent surrogate judges of product owners rather than generic black box graders. First, SAGE uses a small, expert-curated Precedent set rather than large-scale annotation, reducing the annotator heterogeneity problem. Second, SAGE uses absolute graded scores (0-4) rather than pairwise comparisons, because the absolute grades map directly to product decision boundaries for highlighting and filtering results.

To make LLM evaluation practical at scale, knowledge distillation~\cite{hinton2015distilling} compresses expensive teachers into efficient students for relevance tasks under latency and cost constraints~\cite{Vo2024Distillation}. Complementary to model compression, behavioral testing frameworks emphasize maintaining explicit, reviewable expectation suites that function as regression tests over time~\cite{ribeiro2020checklist}. Related principle-driven approaches encode written norms to guide model behavior~\cite{bai2022constitutional}, and practical evaluation guidance stresses end-to-end methodology~\cite{Rudd2025Guide}. SAGE combines distillation for scale with explicit policy specification and traffic-replay simulation to guide AI search iteration.

\section{Problem Formulation and Framework}
\label{sec:framework}
LinkedIn Job Search \cite{juan2025scaling} and People Search \cite{gupta2025retrieval} are large-scale retrieval systems that support natural-language queries with multi-faceted intent such as titles, seniority, and professional relationships.
Members express intents flexibly, often relying on soft constraints and implicit preferences rather than explicit filters. 
Systematically specifying and enforcing a ``ground-truth'' relevance standard that faithfully captures human product judgment is challenging.

We define a search system as a ranking function $f: \mathcal{Q} \times \mathcal{D} \to \mathbb{R}$, where $\mathcal{Q}$ is the space of natural language queries and $\mathcal{D}$ is the document corpus. The ``true'' relevance of a document $d$ to a query $q$ is governed by a latent human utility function $H(q, d)$, which represents ideal human product judgment. In industrial applications, approximating $H(q, d)$ is non-trivial due to the multi-faceted nature of intent. The unbounded semantic space of queries is infeasible for explicit rule enumeration.
For example, for query ``entry level data analyst'', should $H$ assign high utility to a role requiring ``0–2 years of experience'' without explicit ``entry level'' label? How should postings that state ``experience preferred but strong entry-level candidates considered'' be evaluated? Product owners may reasonably disagree on such cases, reflecting legitimate differences in normative judgments.

Without a formal mechanism to resolve these ambiguities, manual evaluation cannot scale, while automated proxies such as engagement metrics provide only indirect and often misleading signals that fail to capture the nuance encoded in $H$. We therefore define the \textbf{Governance Problem} as constructing a Surrogate Judge $\mathcal{J}$ that approximates $H$ by minimizing \textbf{Alignment Divergence} $\Delta(\mathcal{J}, H)$, while simultaneously maximizing \textbf{Coverage} $C$.

\begin{figure*}[t]
    \centering
    \includegraphics[width=\textwidth]{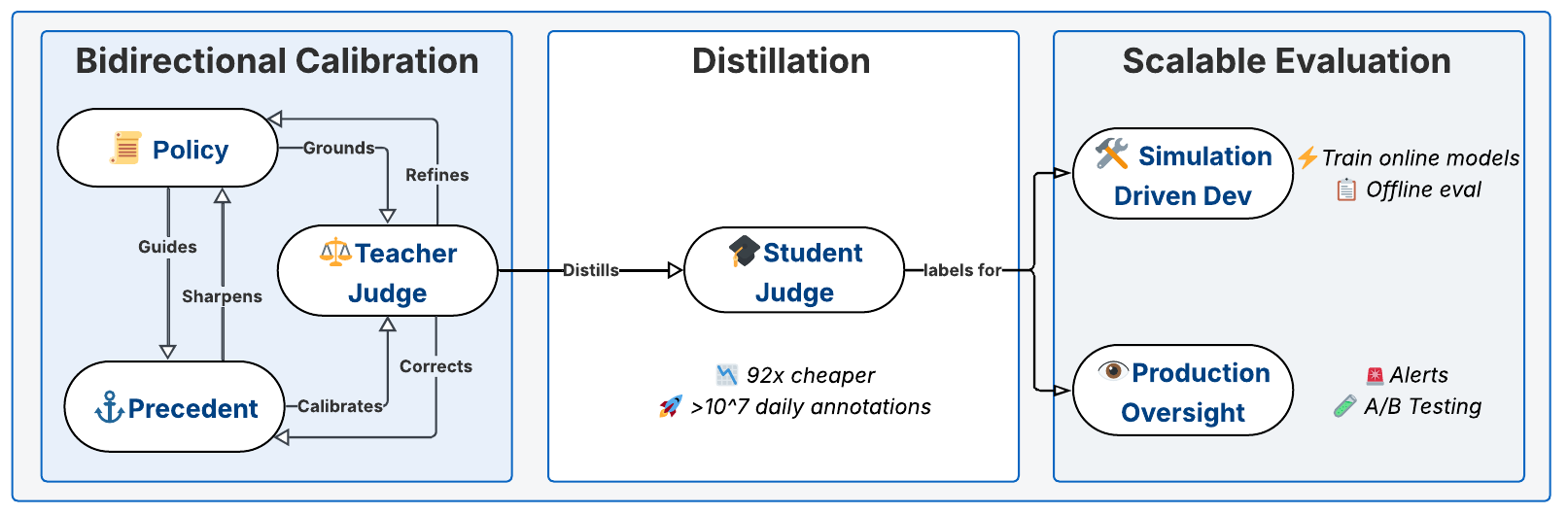}
    \vspace{-16pt}
    \caption{SAGE framework: Bidirectional Calibration produces a calibrated Teacher Judge, which distills into a scalable Student Judge that enables Simulation-Driven Development and Production Oversight.}
    \label{fig:sage-framework}
    \vspace{6pt}
\end{figure*}

We present SAGE (Scalable AI Governance \& Evaluation), a framework designed to make \emph{human-centric evaluation} a commodity for everyday AI search development (Figure~\ref{fig:sage-framework}). SAGE is structured around three core constructs: a natural-language \emph{Policy} ($\mathcal{P}$) that formalizes latent human product judgment into explicit, executable criteria; a \emph{Precedent} ($\mathcal{E}$) that contains a small set of curated canonical judgments that anchor the policy in practice; and an \emph{LLM Surrogate Judge} ($\mathcal{J}$) that is iteratively calibrated against this precedent to execute policy-compliant judgments at scale. This triad enables two essential capabilities: production policy oversight through continuous monitoring of live traffic for regression detection, and simulation-driven development via traffic-replay simulations for training and release gating.

SAGE is built on three foundational design principles that directly address the core challenges of industrial-scale AI governance:
\begin{compact_enum}
    \item \emph{Principle of Bidirectional Calibration}: We model alignment as a dynamic equilibrium where Policy $\mathcal{P}$ guides Precedent $\mathcal{E}$ curation, while Precedent disagreements and Judge $\mathcal{J}$ misalignment expose policy ambiguities. This feedback loop ensures $\mathcal{P}$, $\mathcal{E}$, and $\mathcal{J}$ co-evolve to minimize alignment divergence $\Delta$.
    \item \emph{Principle of Relevance Decomposability and Explainability}: To eliminate ``black-box'' opacity in Judge reasoning, Policy functions as a multi-dimensional rubric decomposing relevance into orthogonal attributes (e.g., Title, Seniority). This enables targeted identification of Judge misalignment during calibration.
    \item \emph{Principle of Comprehensive Coverage}: Achieving the statistical power requisite for oversight ($>10^7$ daily labels) and real-time quality control ($>10^4$ QPS) necessitates economic efficiency. We adopt teacher-student distillation to compress frontier LLM reasoning into a smaller student surrogate while maintaining high agreement.
\end{compact_enum}

The above principles are designed to generalize across search and recommendation domains where explicit queries are present and natural language policy can be tailored to the domain. The Policy-Precedent-Judge triad, bidirectional calibration loop, and relevance decomposability are domain-agnostic and have been applied across multiple LLM-powered search products at LinkedIn.

\section{Calibrating the Teacher Judge}
\label{sec:teacher-judge}

Standard industry practice often treats the Policy as static and the curated labels as immutable ground truth. In production, however, we observed two systematic failure modes: (i) long-tail traffic surfaces novel edge cases that expose underspecified policy, and (ii) consistent LLM application of policy exposes annotation errors even in expert-curated precedent. This motivates us to engineer a system called Bidirectional Calibration, where the Judge does not just learn from human inputs, but actively critiques them, driving updates to both Policy and Precedent.

\subsection{Implementation Decisions \& Trade-offs}
\label{sec:policy-design}
    Our primary design objective is governance utility, prioritizing rapid iteration, interpretability, and actionable diagnostics over maximizing model expressiveness. To this end, we explicitly restricted the Judge’s scope to semantic relevance, decoupling it from mathematical signals like network distance or job posting recency which can be calculated deterministically.
    
    This separation is architectural, not just convenient. We separate semantic        
    relevance from member preferences and business heuristics: the Judge acts as a neutral auditor, strictly measuring the semantic match between query and document, while other system components optimize for ``taste'' (e.g., balancing extended network reach against first degree connections for a people search). By decoupling these concerns, we reserve LLM inference strictly for resolving linguistic ambiguity. This focused scope is further supported by three key architectural decisions:
    
    \textbf{Decision 1: Decomposed Relevance Attributes over Monolithic Evaluation.} Rather than relying on a single monolithic score, we decompose relevance into orthogonal attributes (e.g., Title, Company, Location). We score each attribute independently, then derive the final relevance judgment using weighted heuristics that prioritize the factors most critical to user satisfaction.
   
    Trade-off: This approach constrains the Judge's flexibility to perform loose, context-dependent matching in exchange for granular control.  This composite approach allows us to align the final score with human intuition by strictly prioritizing the signals users care about most, for instance ensuring that a match on Job Responsibilities outweighs a match on Company Size (where users are typically more flexible). Beyond alignment, this also creates a deterministic audit trail, enabling precise attribution of why a specific score was assigned and improving both debuggability and tunability.
    
\textbf{Decision 2: Graded Relevance over Preference Ranking. }Instead of the pairwise ranking (e.g., Bradley-Terry formulations~\cite{hunter2004mm}) common in academic literature, we implemented a graded 5-point scale (0–4). This anchors judgments to concrete member experiences, empowering product owners to iterate on rubrics that explicitly balance inventory liquidity (showing enough results) against precision (showing only the best results) based on the desired user experience.

Trade-off: We prioritize absolute inventory control over the relative ordering advantages of pairwise models. While pairwise ranking identifies the ``best'' option even among poor choices, a graded scale enables the enforcement of explicit quality thresholds: Grades 3–4 drive primary inventory, while Grades 0–1 trigger automatic suppression, converting abstract ``quality'' scores into explicit outcomes that directly enforce the desired user experience.    
    
    \textbf{Decision 3: High-Fidelity Expert Curation over Large-Scale Annotation.} We chose to build a smaller, curated "Precedent" set of a few hundred representative examples annotated by domain experts within the product team rather than relying on large-scale labeling by generalist annotators~\cite{zhou2023lima}. This decision simultaneously optimizes for both precision and velocity: expert annotators produce higher-fidelity labels by rapidly converging on shared interpretations of search intent and explicit liquidity vs. precision trade-offs, while the compact dataset size enables rapid iteration as critical new patterns emerge.
    
    Trade-off: We intentionally accepted a smaller, expert-curated Precedent set to prioritize label accuracy over quantity, thereby enabling policy agility.
    A compact, expert-maintained precedent enables rapid realignment: after a policy revision, the Precedent can be updated within hours, keeping the model synchronized with evolving product judgment.
    
\subsection{The Bidirectional Calibration Loop}
\label{sec:human-calibration}
    The structural agility of SAGE is fully realized in the Bidirectional Calibration loop. By engineering a flexible architecture where information can travel upstream (from Judge to Policy) just as easily as it flows downstream (from Human to Judge), we transform the evaluation process from a rigid waterfall into a responsive system. This ensures that while human expertise sets the initial standard, the Teacher itself can identify opportunities for improvement and drive updates. The components evolve simultaneously through four feedback vectors:

\textbf{    Human $\rightarrow$ Policy (Policy Intuition Gaps):} Expert annotators are explicitly instructed to flag instances where the Policy dictates a grade that contradicts their domain intuition. These domain experts and members of the product team then regularly meet for a Policy Review. These disagreements are then debated with the resolution codified into the continuously updated Policy.

\textbf{    Human $\rightarrow$ Human (Policy Ambiguity Detection):} We utilize inter-rater agreement (measured via Cohen's kappa) as a diagnostic signal for Policy ambiguity. Consistently low agreement on specific query patterns indicates that the Policy fails to address a specific nuance. Disagreements are then evaluated to determine whether they stem from evaluator misapplication of the policy, or from genuine policy under-specification that warrants an explicit policy update. 

\textbf{    Judge $\rightarrow$ Precedent (Adversarial Audit):} We recognize that Precedent is susceptible to human error. When the Judge disagrees with a human label, we audit the reasoning trace to understand the model's decision and confirm whether the policy was applied accurately. We found that the Judge frequently correctly identified evidence buried in verbose documents (e.g., detailed job postings or lengthy member profiles) that human reviewers overlooked. Upon expert validation, we update the Precedent, leveraging the Judge to surface potential errors for human correction.

\textbf{Judge $\rightarrow$ Policy (Edge-Case Discovery):} When the Judge introduces unsupported constraints or applies inconsistent logic, it is often a symptom of policy under-specification. By auditing these reasoning failures, we can precisely identify where the policy fails to cover specific scenarios.

Illustrative toy examples of each of the four feedback vectors operating in production are provided in Appendix~\ref{appendix:toy-calibration}; representative slices of the grading Policy and examples of the Teacher Judge applying it are provided in Appendix~\ref{appendix:grading-examples}. Table~\ref{tab:judge-calibration} shows the efficacy of these feedback vectors as Judge-Precedent agreement improves with each intervention. The Baseline row corresponds to a G-Eval~\cite{liu2023geval}-style setup with GPT-o3 prompted to decompose relevance into attributes in a CoT manner before aggregating into a final score, reaching $\kappa\sim0.67$ prior to bidirectional calibration. For Semantic People Search, through several iterations of bidirectional calibration, we achieved Teacher Judge–Human agreement of 0.77 (linear weighted Cohen's kappa), surpassing the 0.7 threshold for substantial agreement \cite{landis1977measurement} and approaching the practical agreement ceiling of expert product owners (0.83; Figure \ref{fig:linear-kappa-comparison}) in a complex and often subjective domain, establishing a stable foundation for distillation.
\begin{table}[tb]
\centering
\small
\caption{Impact of representative interventions on Judge calibration for Semantic People Search.}
\label{tab:judge-calibration}
\begin{tabular}{lll>{\raggedright\arraybackslash}p{1.25in}}
\toprule
 \textbf{Stage} &\textbf{Kappa}& \textbf{Feedback Vector}&\textbf{Intervention}  \\
\midrule
 Baseline &$0.67$& —&G-Eval-style: GPT-o3 with initial policy\\
 Iteration 1&$0.71$ & Judge→Precedent&Corrected human labeling errors identified by Judge disagreement\\
 Iteration 2&$0.73$ & Human→Policy&Refined "Person Name" policy after annotator disagreement\\
 Iteration 3&$0.75$ & Judge→Policy&Aligned Industry scoring after Judge hallucinated user-declared signals\\
 Iteration 4&$0.77$ & Human→Policy&Standardized scoring for "non-human" profiles\\

\bottomrule
\end{tabular}
\end{table}

\section{Distilling a Production-Scale Student Judge}
\label{sec:student-judge}

While frontier LLMs (e.g., GPT-o3) can act as high-fidelity surrogate judges when calibrated against expert human precedent, they are impractical as day-to-day evaluators for large-scale semantic search. Their inference cost and latency make it prohibitive to continuously run large traffic-replay simulations for iteration and launch gating. This creates an evaluation bottleneck: system owners can generate many candidate variants, but cannot afford to measure relevance fast enough or at sufficient scale to confidently make decisions.

To overcome this constraint, we distill our calibrated Teacher Judge (GPT-o3) into an 8B parameter Student Judge optimized for production throughput while preserving judgment quality. The Student Judge enables tens of millions of daily graded evaluations, turning evaluation into a scalable infrastructure primitive rather than a scarce, expensive resource.

\subsection{Success criteria: Student–Human Alignment}
Our core objective is human-centric evaluation at scale. Accordingly, our primary success metric is Student–Human alignment on the Precedent, measured by linear weighted Cohen’s kappa which accounts for distance-aware disagreement under ordinal 0–4 relevance grades. We target \textbf{linear kappa $\geq$ 0.7}, an industry-standard threshold for substantial agreement and production reliability \cite{landis1977measurement}.

To ensure decision-critical quality beyond aggregate linear kappa, we additionally compute:

\begin{compact_enum}
\item \textbf{F1-Good+}: precision–recall performance for identifying high-quality results ({3,4})

\item \textbf{F1-Poor-}: precision–recall performance for identifying poor results ({0,1})
\end{compact_enum}

Together, these metrics validate that the Student Judge preserves both positive precision and rejection rigor -- properties required for launch gating and regression detection in production.

\subsection{Distillation at Scale}
A key engineering contribution of SAGE is enabling a calibrated Teacher Judge to operate at production scale through teacher–student distillation. We fine-tune the Student Judge on an open-source decoder-only \textbf{8B} backbone, chosen for strong instruction following and a favorable latency–quality tradeoff. To ensure behavioral consistency under distillation, the Student Judge \textit{reuses the same prompt structure and output schema as the Teacher Judge}.

We construct Student Judge training data by applying the Teacher Judge to large-scale stratified samples of production traffic and aspirational queries - natural-language, intent-rich queries that the semantic search enables beyond keyword-style search (e.g., ``Who can help me with mortgage in the Bay Area'').  To prevent collapse toward high-relevance predictions, we rebalance underrepresented score classes (0--3) toward a near-uniform distribution, producing a training corpus of 312K examples.

We train the Student Judge via supervised instruction tuning~\cite{wei2021finetuned} over hydrated prompts paired with Teacher Judge outputs. While early iterations employed parameter-efficient LoRA \cite{hu2021lora}, we found that full-parameter fine-tuning delivered significantly higher agreement stability for graded judgments. 
The final production Student therefore adopts full-parameter fine-tuning with ZeRO-3–style optimizer~\cite{rajbhandari2020zero}. By sharding model parameters, gradients, and optimizer states across workers, this approach ensures memory-efficient training at scale.

We normalize Student Judge cost = 1.0 and report all other evaluation costs relative to this baseline in Table \ref{tab:cost_comparison}. Human judge costs are estimated using Amazon Mechanical Turk pricing \cite{amt_pricing}, assuming an absolute lower bound on compensation with no qualification requirements; as such, this estimate represents a conservative floor on human evaluation cost. Even under this optimistic assumption, human evaluation is approximately $154\times$ more expensive than the Student Judge, while the calibrated Teacher Judge remains $\sim92\times$ more costly. These results highlight the extent to which distillation enables scalable, high-throughput evaluation that would be economically impractical with direct human labeling.
\begin{table}[tb]
\centering
\caption{Relative evaluation cost comparison (normalized to Student Judge = 1.0).}
\label{tab:cost_comparison}
\begin{tabular}{l c}
\hline
\textbf{Judge Type} & \textbf{Relative Cost} \\
\hline
Minimum Lower-Bound Human Judge & 154$\times$ \\
Teacher Judge                  & 92$\times$  \\
Student Judge                  & 1$\times$   \\
\hline
\end{tabular}
\end{table}

\subsection{Experiments: Optimizing the Student Model}
\begin{figure}[tb]
    \centering
\hspace{-2pt}    
\includegraphics[width=\linewidth]{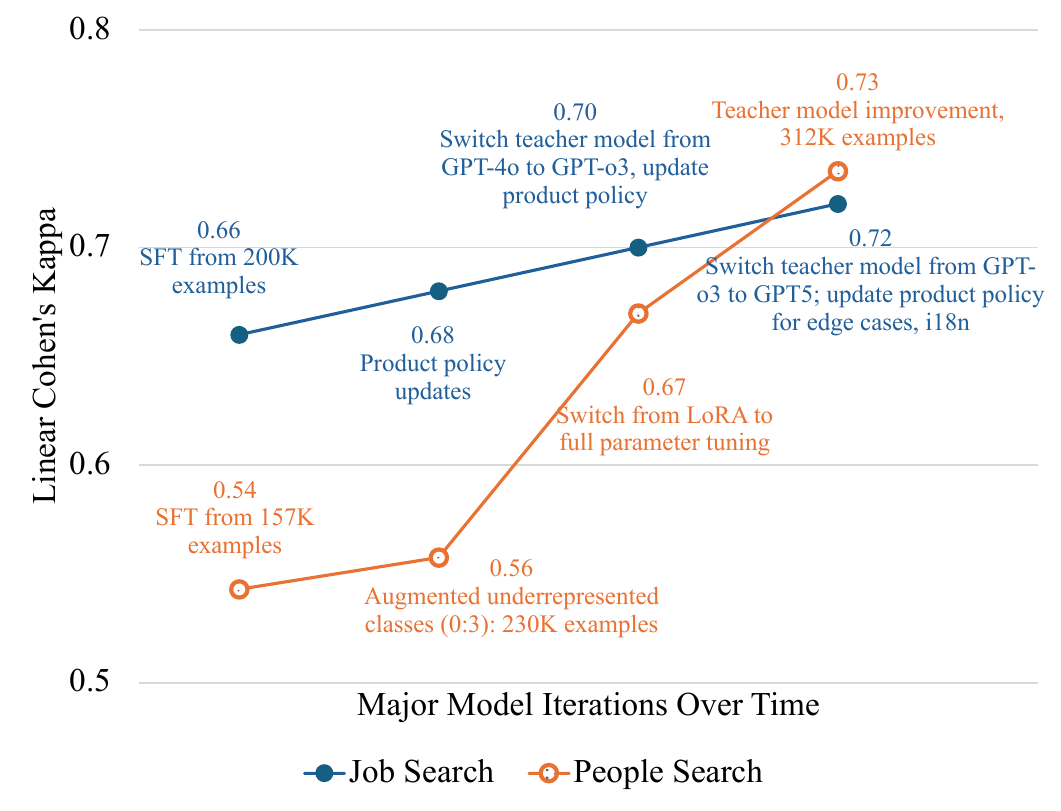}
    \caption{Linear Cohen's kappa of Student Judge across major training iterations for Job Search and People Search.}
    \label{fig:student-judge-iteration}
\end{figure}

We iteratively improved the Student Judge through a sequence of major refinements applied across both \textbf{Job Search} and \textbf{People Search} by optimizing training techniques. Figure \ref{fig:student-judge-iteration} summarizes the key milestones, including augmenting underrepresented score classes (0–3), switching from LoRA to full fine-tuning, and Teacher model improvement. As a result, Student–Human alignment in both domains surpasses the \textbf{0.7 linear kappa} threshold for substantial agreement, reaching \textbf{0.72} for Job Search and \textbf{0.73} for People Search. Additional experiment details for People Search including F1 Good+/F1 Poor- metrics improvements are shown in Table~\ref{tab:sage_iterations}.

\begin{figure}[tb]
    \vspace{6pt}
    \centering
    \includegraphics[width=\linewidth]{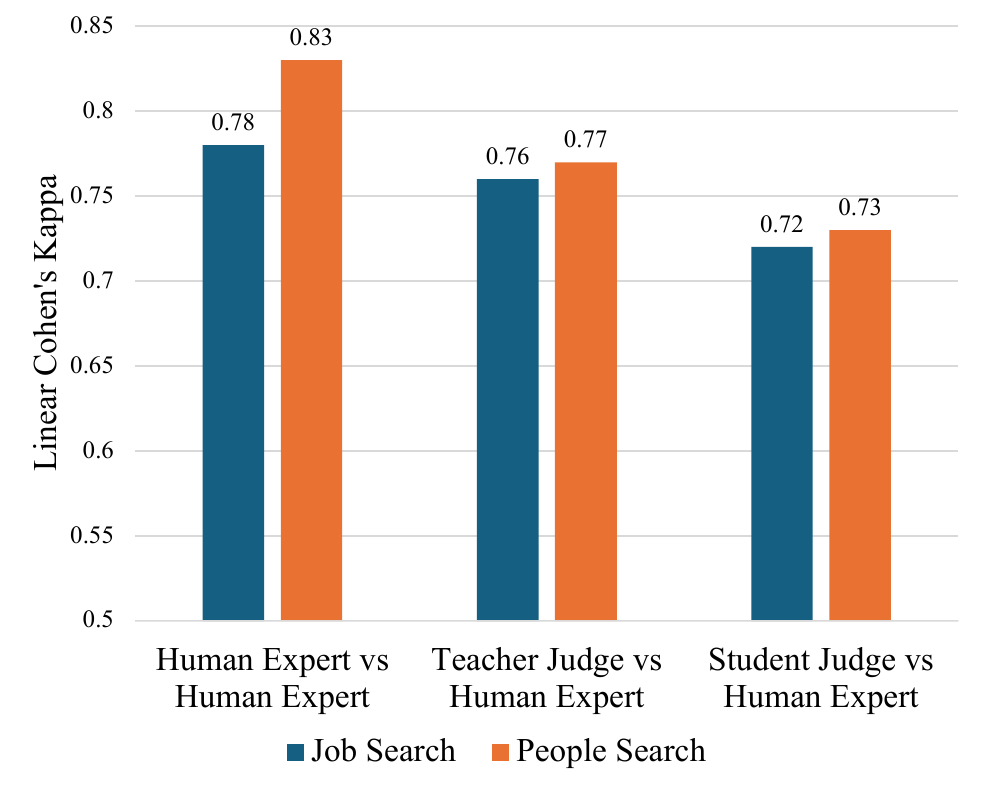}
    \caption{Linear Cohen's kappa scores across different judge comparisons.}
    \label{fig:linear-kappa-comparison}
\end{figure}

Figure \ref{fig:linear-kappa-comparison} further contextualizes alignment ceilings and distillation fidelity. \textbf{Human–Human agreement (0.78 for Job Search; 0.83 for People Search)} represents the \emph{practical ceiling} of this task: even expert annotators do not fully agree, so no judge should be expected to exceed this limit in a meaningful way. The \textbf{Teacher–Human alignment (0.76 for Job Search; 0.77 for People Search)} forms the effective upper bound for distillation: since the Student is trained to imitate the Teacher, the Student–Human alignment is naturally constrained by how well the Teacher itself matches human precedent. In this setting, the Student achieves \textbf{0.72 for Job Search; 0.73 for People Search}, approaching the Teacher’s ceiling while operating at production scale. Additionally, as an illustrative example in People Search, Student-Teacher agreement reaches 0.81, indicating faithful transfer of reasoning logic.

\begin{table}[tb]
\centering
\caption{Impact of training strategies on People SAGE Student Judge performance.}
\label{tab:sage_iterations}
\footnotesize
\setlength{\tabcolsep}{4pt}
\renewcommand{\arraystretch}{1.15}
\begin{tabularx}{\columnwidth}{@{}Xcccc@{}}
\toprule
\textbf{Major Model Iteration} &
\textbf{Linear $\kappa$} &
\textbf{Quadratic $\kappa$} &
\textbf{F1 Good+} &
\textbf{F1 Poor-} \\
\midrule
Initial baseline: 157K training data. &
0.5429 & 0.6465 & 0.8599 & 0.6139 \\
Augmented underrepresented score classes (0--3); 230K training data. &
0.5576 & 0.6774 & 0.8562 & 0.6629 \\
Switched from LoRA to full fine-tuning. &
0.6697 & 0.7984 & 0.8884 & 0.8148 \\
Teacher improvement; 312K training data. &
\textbf{0.7351} & \textbf{0.8289} & \textbf{0.9105} & \textbf{0.8459} \\
\bottomrule
\end{tabularx}
\end{table}

\section{Leveraging the Student Judge}
\label{sec:use-cases}

\begin{figure}[tb]
    \centering
    \includegraphics[width=\columnwidth]{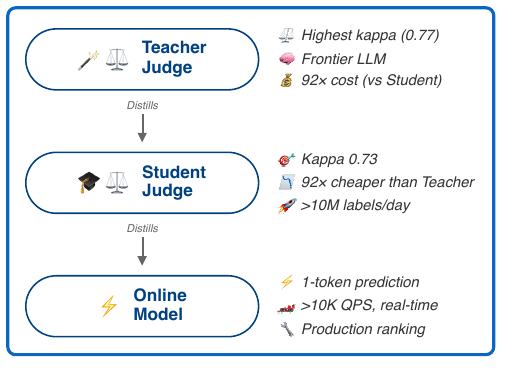}
    \vspace{-22pt}
    \caption{Our distillation cascade: a frontier Teacher Judge is distilled into a scalable Student Judge, which is further compressed into an Online Model for production ranking.}
    \label{fig:distillation-overview}
\end{figure}

We strategically integrate the Student Judge annotations into our system development lifecycle in a few key ways. First, we distill the Student Judge reasoning into an ultra-compact architecture that acts as a real-time policy enforcement mechanism (Figure~\ref{fig:distillation-overview}). Next, we define a holistic metric-framework from Student Judge annotations with which to quantify system performance at scale. These relevance metrics then help expedite and improve business decisions during offline candidate selection, A/B testing, and continuous production monitoring for regression detection.

By implementing the solutions above, we highlight two recent online experiments that collectively realized (+0.25\%) in LinkedIn Daily Active Users (DAU), the largest step-function change in growth the LinkedIn Job Search product has driven in years. 

\subsection{Distillation for Online Serving}
To operationalize quality control within the latency-critical path of real-time search ($>10^4$ QPS) we had to further distill the Student Judge reasoning capabilities into an ultra-compact architecture optimized for high-throughput scoring. We managed this by capitalizing on a few techniques \cite{borisyuk2026semantic}:

\begin{compact_enum}

\item \textbf{Structural Pruning.} To excise redundant parameters, we employed structural pruning on a 0.6B-parameter language model to achieve a \verb|~|40\% reduction in model size.

\item \textbf{Reasoning Internalization.} To obviate the latency of autoregressive Chain-of-Thought generation \cite{wei2022chain}, we train the model to encode policy logic directly into its dense internal representations. By supervising the model with high-precision rationales during training, we enable it to bypass intermediate decoding steps at inference time, collapsing complex reasoning into a direct, low-latency 1-token prediction.

\item \textbf{Context Minimization.} To mitigate the quadratic cost of self-attention, we minimize input context via an offline summarization pipeline. We employ a specialized summarizer~\cite{arora2026high} trained with a retention objective that preserves semantic evidence while aggressively reducing token length. Offline ablations show that this compression substantially reduces sequence length with minimal loss in ranking fidelity.

\end{compact_enum}

Collectively, these optimizations yield the \textbf{Online Model}, a high-fidelity ranking gatekeeper that operates within strict millisecond-level SLAs. The Online Model achieves a linear weighted kappa of 0.767 against the 8B Student Judge. When treating Student Judge labels as ground truth, the model further achieves an NDCG@5 of 0.968, indicating that its ranking behavior closely matches that of the Student Judge.

\subsection{LLM-Based Relevance Metrics}

Using Student Judge annotations, we define a metric framework with which to evaluate offline candidates and online production traffic. This framework measures how effective our system is at three tasks:

\begin{compact_enum}
\item \textbf{Recall}: Find all of the relevant documents

\item \textbf{Precision}: Find \textit{only} the relevant documents

\item \textbf{Ranking}: Present documents from most to least relevant
\end{compact_enum}

For operational purposes, we classify a document as a ``Poor Match'' if its relevance score $s \leq 1$, and a ``Good Match'' if $s \geq 3$. Suppose we aim to serve the top $K$ out of $N$ documents scored and $G$ total "Good" documents retrieved. Then, for each document, $i$, the mental framework above can be formalized as:

$$
\text{Good Recall (GR) @ K} = \frac{1}{min(K, G)}\sum_{i=1}^{min(K, N)} \mathds{1}_{\{s \geq 3\}}
$$

$$
\text{Poor Match Rate (PMR) @ K} = \frac{1}{min(K, N)}\sum_{i=1}^{min(K, N)} \mathds{1}_{\{s \leq 1\}}
$$

$$
\text{NDCG @ K} = \sum_{i=1}^{min(K, N)}\frac{2^{s_i} - 1}{\text{IDCG} \cdot \log_2(i + 1)}
$$

where NDCG and IDCG are the Normalized and Ideal Discounted Cumulative Gain respectively~\cite{jarvelin2002cumulated}. 

The parameter $K$ is chosen to be the number of documents expected to be impressed to the member, such that the metric is sensitive to changes that ultimately impact users.

\subsection{Offline Candidate Selection}
One foundational step of AI search system development is the curation and scrutiny of up to hundreds of model candidates prior to online testing. We have the Student Judge annotate samples of counterfactual data, with which to vet and gate models that do not meet our quality standard. 

This process consists of the following stages:

\begin{enumerate}
    \item \textbf{Sampling}: We sample real-user queries from historical production traffic, optionally up-weighting less frequent examples to measure model performance on a high-coverage, representative data set.
    \item \textbf{Simulation}: We pre-deploy candidate models and replay the sampled queries against their online endpoints. Because the Student Judge evaluates relevance directly from query-document pairs, we can score outputs immediately without waiting for user engagement signals to accumulate. 
    \item \textbf{Evaluation}: We leverage our Student Judge to derive annotation-based metrics to compare model candidates along important query \& user dimensions.  
\end{enumerate}

Adopting this process led to quantifiable improvement in \textbf{proactive regression detection} and \textbf{iteration velocity}. For example, upgrading from manual spot checks to aggregate metrics empowered us to catch a -16\% GR@10 regression on company-queries before exposing the variant to users. By circumventing the need to collect user data, we reduced the candidate selection cycle duration from 2 weeks to just 3 days (80\% reduction). 

\subsection{Online Experimentation}

\begin{figure}[tb]
    \centering
    \includegraphics[width=\columnwidth]{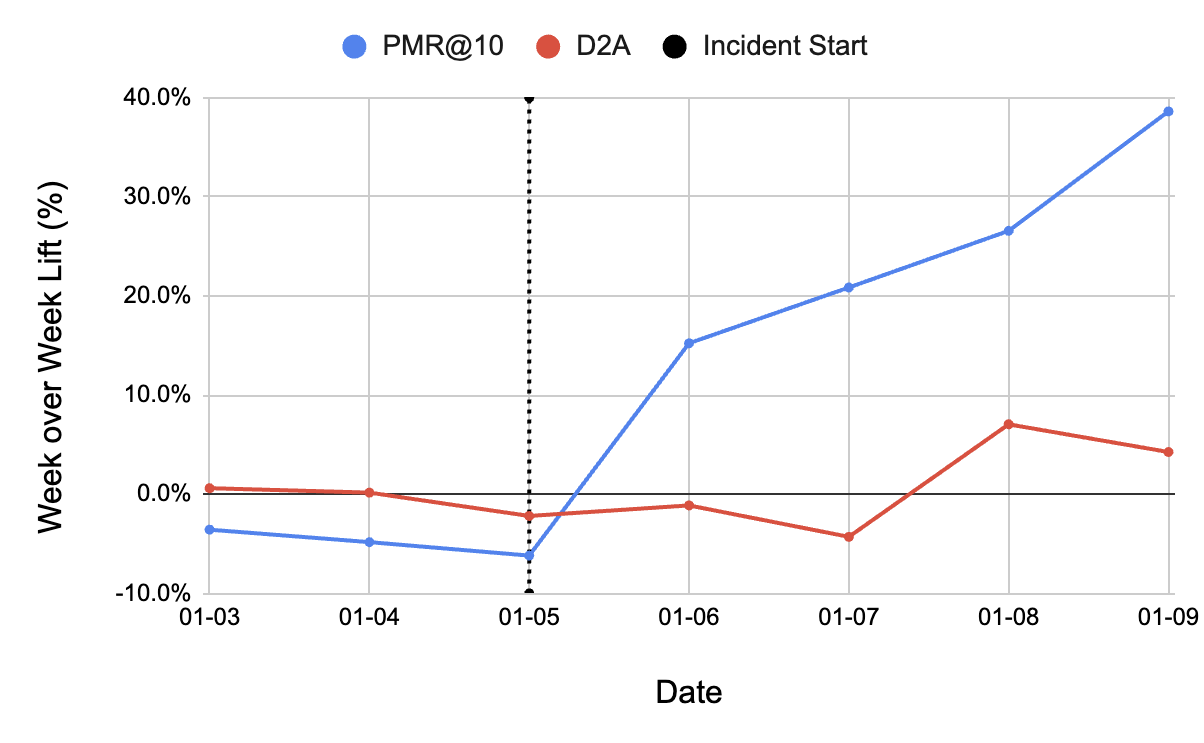}
    \vspace{-24pt}
    \caption{PMR@10 vs. Dismiss to Apply Ratio (D2A) in the presence of a production incident degrading search quality.}
    \label{fig:pmr-regression}
\end{figure}

Ramp candidates that pass through the offline selection process are then deployed online in an A/B test configuration to measure the causal impact to the product ecosystem. As with offline evaluation, we leverage the Student Judge to derive policy-based relevance metrics from experiment data, with which to choose a winning variant. The overall process is similar (sample, score, scrutinize); however, there are a few key implementation differences: 

\begin{enumerate}

\item \textbf{Data Generation Process}: While offline evaluation requires simulating counterfactual data from a historical snapshot, online experiments provide us with real production traffic authentic to live user behavior. To ensure we capture organic shifts in the data distribution, we set up a recurring, daily sampling pipeline capable of annotating over 100 million query/document pairs in a given week.

\item \textbf{Sampling Strategy}:
We choose to stratify-sample along different dimensions for offline candidate selection vs. A/B experiment decisions. For offline candidate selection, the focus is typically on upsampling and improving system performance on long-tail queries that are perhaps less common (e.g., foreign language). Meanwhile, during online experimentation the primary focus is high-resolution measurement across all of the major product channels/surfaces (e.g., email, jobs home, feed). This not only reflects how we structure development of our system, but also ensures that excessively high-volume channels don't bias our understanding of quality throughout the ecosystem. 

\end{enumerate}

Once a winning variant is chosen and ramped as the Majority Member Experience (MME), we continue to surface policy-based relevance metrics derived from the daily pipeline in operational dashboards to monitor system health on an ongoing basis. This enables us to catch system regressions that don't immediately manifest as declines in engagement. For example, Figure~\ref{fig:pmr-regression} shows a production incident where search quality degraded, causing a \textbf{40\% week-over-week increase in PMR@10}, yet our closest engagement-based proxy for relevance, Dismiss to Apply Ratio (D2A), did not meaningfully deviate from neutral.   

\subsection{Overall Business Impact}

We evaluate the end-to-end impact of SAGE through two sequential online A/B experiments. In both deployments, we integrated the low-latency Online Model into the candidate generation pipeline for LinkedIn’s highest-volume Job Search Alert campaign.

Table~\ref{tab:online_exp} reports the impact in terms of both the overall quality of the content served, and the subsequent Top-of-Funnel (TOF) growth (represented by DAU). All reported metrics are statistically significant ($p$<0.05).

\begin{table}[tb]
\centering
\caption{Online Experiment Results: Job Search Alert Campaign}
\label{tab:online_exp}
\begin{tabular}{lcccc}
\toprule
\textbf{} & \textbf{PMR@10} & \textbf{NDCG@10} & \textbf{GR@10} &  \textbf{DAU} \\
\midrule
Iteration 1  & -28.9\% & +8.43\%  & +10.1\% & +0.12\% \\
Iteration 2  & -50.9\% &  +7.83\%  & +16.5\% & +0.13\% \\
\bottomrule
\end{tabular}
\end{table}

We observed that enforcing a policy-based relevance bar yielded a significant reduction in PMR@10, effectively suppressing low-quality content. This improvement was accompanied by gains in NDCG@10 and Good Recall (GR@10), suggesting that the pipeline both demoted poor matches and increased exposure of high-quality opportunities in top positions.

Most importantly, we observed that policy-based relevance improvements occur alongside substantial lifts in DAU. 
This increase is driven by two complementary mechanisms enabled by improved policy-grounded relevance: (i) higher effective recall of good opportunities, allowing increased campaign volume without degrading quality, and (ii) reduced exposure to poor matches, improving member trust and downstream engagement.

\section{Conclusion}
\label{sec:conclusion}

\subsection{Summary of Results}
We presented \textbf{SAGE} (Scalable AI Governance \& Evaluation), an LLM-based evaluation framework built on three principles: \emph{bidirectional calibration} between \textbf{Policy}, \textbf{Precedent}, and \textbf{Judge}; \emph{relevance decomposability} for interpretable failure attribution; and \emph{comprehensive coverage} enabled through teacher-student distillation.

Through iterative bidirectional calibration, the Teacher Judge achieved a {0.77} linear weighted Cohen’s kappa against expert human judgments, matching the inter-rater reliability of expert product owners (Section~\ref{sec:teacher-judge}). Distilling the Teacher into a production-scale Student Judge preserved alignment quality: the Student achieved human agreement of {0.72} (Job Search) and {0.73} (People Search), both exceeding the 0.7 threshold for substantial agreement (Section~\ref{sec:student-judge}), at {92$\times$} lower cost than the Teacher.

Deployed to LinkedIn search products, SAGE guided model iteration through simulation-driven development, distilling online models and accelerating their offline evaluations. In production, it powered policy oversight that measured ramped model variants and detected regressions invisible to engagement metrics. Collectively these capabilities drove a {+0.25\%} lift in LinkedIn daily active users (Section~\ref{sec:use-cases}).

\subsection{Lessons Learned}

\noindent\textbf{High-Fidelity Curation Beats High-Volume Annotation.}
The limiting factor in actionable evaluation was not model capacity but the availability of high-quality, policy-consistent precedents. In practice, a small set of carefully curated canonical cases provided more leverage than large-scale labeling over an unbounded query space. As semantic systems scale, the most valuable human effort shifts from \textit{producing labels} to \textit{resolving ambiguity and codifying norms}.

\vspace{6pt}
\noindent\textbf{The Judge Can Audit the Ground Truth.}
Contrary to the assumption that human labels are immutable, we found that a calibrated Judge frequently surfaced overlooked evidence in verbose documents and exposed inconsistencies in ``golden'' datasets. Treating model disagreements as adversarial audits enabled systematic correction of precedent errors and accelerated policy refinement. This suggests that LLM-based evaluation can improve not only model quality but also the quality of the supervision itself.

\vspace{6pt}
\noindent\textbf{Policy Must Be a Versioned Artifact.}
As policy definitions evolve, evaluation metrics can shift even when the underlying retrieval system is unchanged. Without explicit policy versioning, such shifts appear as false regressions and erode stakeholder trust. Treating Policy as a versioned software specification allowed us to attribute metric deltas to definitional changes rather than model failures, enabling stable long-term governance.

\vspace{6pt}
\noindent\textbf{Decomposition Enables Debuggability and Adoption.}
Scalar relevance scores are insufficient for diagnosing failures in complex retrieval stacks. Decomposing relevance into interpretable attributes made regressions actionable, facilitated faster iteration cycles, and improved cross-functional adoption by providing a shared vocabulary for discussing product judgment.

\vspace{6pt}
\noindent\textbf{Semantic Evaluation Complements Behavioral Proxies.} While engagement signals (e.g., clicks) remain the industry standard for monitoring, they are susceptible to popularity bias and often fail to surface ``silent'' relevance degradation. In contrast, SAGE shows that a calibrated Judge can systematically detect quality regressions that are latent in behavioral metrics. Moreover, the observed association between offline policy-based relevance improvements and online DAU lift suggests that operationalized product judgment provides a more stable and diagnostic indicator of long-term ecosystem health than raw interaction statistics alone.

\begin{acks}
We thank our Product and Engineering leadership, particularly \textbf{Mike Jennings}, \textbf{Diane Jen-mei Lee}, \textbf{Luke Simon}, \textbf{Rohan Rajiv}, and \textbf{Adi Pruthi}, who championed SAGE from its inception. We also express special appreciation to the LLM serving team, the Seeker Engineering team, and the many talented engineers across LinkedIn whose dedication made this work possible.
\end{acks}


\balance
\bibliographystyle{ACM-Reference-Format}
\bibliography{illuminator_bib}

\appendix

\section{Toy Examples of Bidirectional Calibration}
\label{appendix:toy-calibration}
We provide one illustrative example of each of the four feedback vectors operating in production, two drawn from LinkedIn Job Search and two from LinkedIn People Search. These examples demonstrate how Policy and Precedent co-evolve through the bidirectional calibration loop described in Section~\ref{sec:human-calibration}.

\begin{table}[!htbp]
\centering
\small
\caption{Illustrative examples of feedback vectors driving iterative alignment in Job Search and People Search.}
\label{tab:appendix-toy-calibration}
\renewcommand{\arraystretch}{1.2}
\begin{tabularx}{\columnwidth}{@{}lX@{}}
\toprule
\textbf{Vector} & \textbf{Toy Example} \\
\midrule
Human $\rightarrow$ Policy &
In early iterations of Job Search, PM experts graded ``title'' relevance based on whether the job posting contained an exact or partial match of the search query. During annotation, we observed that identical titles can involve different responsibilities and that distinct titles can involve similar responsibilities. We therefore pivoted the Policy to evaluate the overlap of job responsibilities between the search query and the job posting. \\
\midrule
Human $\rightarrow$ Human &
In early iterations of Job Search, PM experts graded ``seniority'' relevance based on band differences between the query and the job posting, but encountered significant peer disagreement. Investigation revealed two structural issues. First, seniority gaps are non-uniform: there are meaningful role differences between IC, first-line manager, and manager-of-manager (e.g., for a ``senior software engineer'' query, a ``staff software engineer'' job is more relevant than a ``software engineering manager'' job). Second, job seekers respond asymmetrically to upward and downward role mismatches, treating downward mismatches as disqualifying. The Policy was updated to factor in both functional and emotional needs, after which peer alignment improved significantly. \\
\midrule
Judge $\rightarrow$ Precedent &
In People Search, the Judge often identified information buried in a member profile that a human annotator would overlook (e.g., a historical position at a company referenced in the query). Such cases prompted Precedent updates after expert validation. \\
\midrule
Judge $\rightarrow$ Policy &
In People Search, the Policy initially directed the Judge to interpret and score ambiguous queries under the ``most favorable'' interpretation, to account for multiple possible meanings (e.g., ``Spencer Construction'' might be a company, or a person named Spencer who works in construction). However, the Judge applied this rule inconsistently and often over-reached, scoring too favorably. The Policy was refined to focus on the dominant interpretation, relying on users to refine queries for rare or edge-case scenarios. \\
\bottomrule
\end{tabularx}
\end{table}

\section{Grading Policy and Judge Reasoning}
\label{appendix:grading-examples}
We show representative slices of the grading Policy: Job Search rubrics for the ``location'' and ``title/function'' attributes (Table~\ref{tab:appendix-grading-policy}), and the People Search rubric for the ``location'' attribute (Table~\ref{tab:appendix-grading-policy-ps}). These rubrics were established through several rounds of Precedent curation and expert discussion. We then show two illustrative examples of the Teacher Judge applying the Policy in production; each example shows the query's decomposition into orthogonal attributes, the rubric slice used, and the Judge's reasoning.

\begin{table}[H]
\centering
\footnotesize
\caption{Representative slice of the Job Search Policy for grading the Location and Title/Function attributes. Each row is a grade level (0--4); columns are the two attributes.}
\label{tab:appendix-grading-policy}
\setlength{\tabcolsep}{3pt}
\renewcommand{\arraystretch}{1.2}
\begin{tabularx}{\columnwidth}{@{}lXX@{}}
\toprule
\textbf{Score} & \textbf{Location (incl.\ radius)} & \textbf{Title / Function} \\
\midrule
4 (Great) & Job location matches (equal or within) the query intent & Job responsibilities exactly match the query intent \\
\midrule
3 (Good) & N/A & Job responsibilities closely match the query intent ($>$70\% skill overlap) \\
\midrule
2 (Fair) & Job location is broader than the query intent, with no entry barrier for seekers & Job responsibilities somewhat match the query intent with possible specialization differences; seeker has a chance ($>$30\% skill overlap, no barrier) \\
\midrule
1 (Poor) & N/A & Job responsibilities minimally match the query intent; seeker has low willingness or high entry barrier \\
\midrule
0 (Unacceptable) & Job location does not match the query intent & Job responsibilities are significantly different from the query intent \\
\bottomrule
\end{tabularx}
\end{table}

\begin{table}[H]
\centering
\small
\caption{Representative slice of the People Search Policy for grading the Location attribute.}
\label{tab:appendix-grading-policy-ps}
\renewcommand{\arraystretch}{1.2}
\begin{tabularx}{\columnwidth}{@{}lX@{}}
\toprule
\textbf{Score} & \textbf{Use When} \\
\midrule
4 (Great) & Exact location match, or the profile location falls directly inside the queried area. \\
\midrule
3 (Good) & Near or equivalent match, such as an adjacent suburb, nearby city in the same metro, or well-known commuter corridor. \\
\midrule
2 (Fair) & Same general region, state, or province, but not the specific requested location. \\
\midrule
1 (Poor) & Only a tenuous geographic connection, such as the same broad part of the world. \\
\midrule
0 (Unacceptable) & No meaningful location match. \\
\bottomrule
\end{tabularx}
\end{table}

\FloatBarrier

Tables~\ref{tab:appendix-judge-example1} and~\ref{tab:appendix-judge-example2} show two illustrative examples of the Teacher Judge applying the Policy in production: one Job Search query and one People Search query. Each example reports the query, the search object, the attributes evaluated, the Judge's reasoning, and the final score.

\begin{table}[H]
\centering
\small
\caption{Example 1: Teacher Judge reasoning on a Job Search query.}
\label{tab:appendix-judge-example1}
\renewcommand{\arraystretch}{1.2}
\begin{tabularx}{\columnwidth}{@{}lX@{}}
\toprule
\textbf{Query} & ``Admin Assistant'' (location: NYC) \\
\midrule
\textbf{Search object} & Title ``Mail Handler''; location header lists NYC but the description requires relocation to Florida; description excerpt: ``Handling mail and needing to stand for 8 hours a day\ldots'' \\
\midrule
\textbf{Attributes evaluated} & Title, Location \\
\midrule
\textbf{Judge reasoning} & Location (0): the posting header lists NYC, but the description states Florida; the job is not stated as remote. Role (1): the query seeks an office-type admin role, which is very different ($\sim$20--30\% overlap). \\
\midrule
\textbf{Final score} & Location = 0; Role = 1; Final = 0 \\
\bottomrule
\end{tabularx}
\end{table}

\begin{table}[H]
\centering
\small
\caption{Example 2: Teacher Judge reasoning on a People Search query.}
\label{tab:appendix-judge-example2}
\renewcommand{\arraystretch}{1.2}
\begin{tabularx}{\columnwidth}{@{}lX@{}}
\toprule
\textbf{Query} & ``James Dutton, Utah'' \\
\midrule
\textbf{Search object} & Member named James Dutton, location Evanston, Wyoming, working at a company based in Park City, UT \\
\midrule
\textbf{Attributes evaluated} & Person Name, Location \\
\midrule
\textbf{Judge reasoning} & Exact name match. Member is based in Evanston, WY (a commuter community on the Utah border, functionally part of the northern Utah/Wasatch labor market) and works in Park City, UT. \\
\midrule
\textbf{Final score} & Person Name = 4; Location = 3; Final = 3 \\
\bottomrule
\end{tabularx}
\end{table}

\end{document}